\documentclass[a4paper]{jpconf}

\newcommand{\Jp}{J/\psi}

\newcommand{\GeV}{\mbox{\,GeV}}
\newcommand{\TeV}{\mbox{\,TeV}}
\newcommand{\mb}{\mbox{\,mb}}

\newcommand{\nb}{\mbox{\,nb}}
\newcommand{\pb}{\mbox{\,pb}}

\usepackage{graphicx}
\begin{document}
\title{Phenomenological analysis of the possible impact of \\
Double Parton Scattering in double $\Jp$ production
at the COMPASS detector using the CERN $\pi^-$ beam at $190 \GeV/c$}

\author{Sergey Koshkarev}

\address{Institute of Physics, University of Tartu, 51010 Tartu, Estonia}

\ead{sergey.koshkarev@ut.ee}

\begin{abstract}
In this paper we discuss the impact of the Double Parton Scattering (DPS)
mechanism for the production of $\Jp$ pairs at the COMPASS energy. We find
that the kinematics suppresses the DPS cross section by a factor of $\sim1/2$.
The upper limit for the double $\Jp$ production DPS cross section at the
COMPASS energy is estimated. The Feynman-$x$ distributions of double $\Jp$
production with the CERN $\pi^-$ beam are presented.\footnote{In this paper 
$x_F$ denotes the Feynman-$x$ in the laboratory frame while $x_F^*$ denotes 
the Feynman-$x$ in the center-of-mass system.}
\end{abstract}

\section{Introduction}

\begin{figure}[h]
\includegraphics[width=14pc]{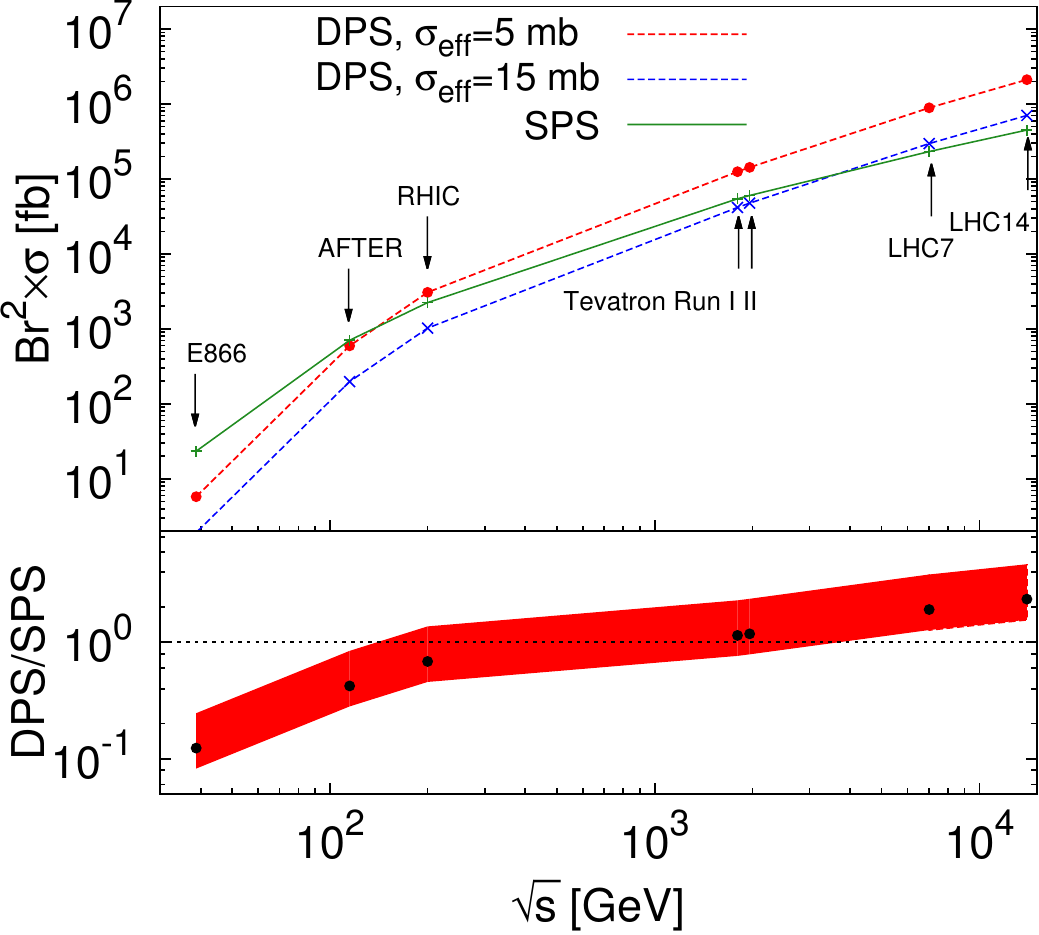}\hspace{2pc}%
\begin{minipage}[b]{14pc}\caption{\label{fig:lansberg_DPS}(Upper panel)
Cross sections of (prompt-)$J/\psi$ pair production via SPS and DPS mechanisms 
for two values of $\sigma_{\rm eff}$ as a function of $\sqrt{s}$. (Lower panel)
DPS over SPS yield ratio for $5 < \sigma_{\rm eff} < 15\mb$. The black circles
correspond to $10\mb$ [Aside from the choice of $\sigma_{\rm eff}$, no
theoretical uncertainties are included~\cite{Lansberg:2015lva}].}
\end{minipage}
\end{figure}

The significance of the double parton scattering (DPS) in associate charmonium 
production has been investigated by the Tevatron and the LHC by measuring the
productions of $\Jp+W$~\cite{Aad:2014rua}, $\Jp+Z$~\cite{Aad:2014kba},
$\Jp + \Upsilon$~\cite{Abazov:2015fbl}, $\Jp$+charm~\cite{Aaij:2012dz}, 
and $\Jp+\Jp$~\cite{Aaij:2011yc,Abazov:2014qba,Khachatryan:2014iia,%
Aaboud:2016fzt,Aaij:2016bqq}. Even though we don't expect DPS as the leading
production mechanism at the COMPASS energy ($\sqrt{s} \approx 19\GeV$), this
contribution is expected to be far from zero (cf.\ the lower panel in
Fig.~\ref{fig:lansberg_DPS})~\cite{Lansberg:2015lva}. Moreover, based on the
NA3 data on double $\Jp$ production using the CERN $\pi^-$ beam at $150$ and
$280\GeV/c$ with incident on a platinum target~\cite{Badier:1982ae} we can
expect up to 100 double $\Jp$ events at COMPASS~\cite{Aghasyan:2017jop,%
Riedl:2018}. Therefore and for many other reasons it is interesting to
investigate a possible contribution of such a mechanism to double $\Jp$
production at the COMPASS detector.

\section{Double $\Jp$ production at the COMPASS energy}

COMPASS is a fixed target experiment utilizing the high intensity $\pi^-$
beam of $190\GeV/c$ at the Super Proton Synchrotron at CERN for Drell--Yan
(DY) measurements to produce charmonium, possible exotic states and dimuons  
with incidents on Ammonia (${\rm NH_3}$), Aluminium (Al) and 
Tungsten (W) targets~\cite{Abbon:2014aex}. The experiment had several DY runs
in 2014, 2015 and 2018. The COMPASS DY configuration setup is quite similar to
the NA3 setup.

\subsection{Double Parton Scattering}

In case of hadron--nucleus collisions the general formula for the predicted
DPS cross section for $\Jp$ pairs is given by~\cite{dEnterria:2014mzh}
\begin{equation}\label{eq:DPS_general}
\sigma_{\rm DPS}^{h A}(\Jp\Jp)= \frac{1}{2}
\frac{\sigma(\Jp)^{h N}\sigma(\Jp)^{h N}}{\sigma_{\rm eff}^{h A}} \, ,
\end{equation}
where $\sigma(\Jp)^{h N}$ denotes the single $\Jp$ hadron--nucleon cross
section and $\sigma_{\rm eff}^{h A}$ is the effective hadron--nucleus DPS
cross section.

Let us remind the reader that in Eq.~(\ref{eq:DPS_general}) the production of
each $\Jp$ in hadron--nucleon collisions is assumed to be an independent
process. However, it is easy to see that the production threshold of the $\Jp$
pair is already more than $30\%$ of the COMPASS energy (cf.\
Tab.~\ref{tab:energy_scale}). Therefore, we cannot assume the production of
charmonium states as independent processes. 
\begin{center}
\begin{table}[h]
\caption{\label{tab:energy_scale} Energy scales for Super Proton Synchrotron,
Tevatron and LHC accelerators}
\centering
\begin{tabular}{@{}*{7}{l}}
\br
Accelerator&Energy ($\sqrt{s}$)&Colliding Mode\\
\mr
\verb"Super Proton Synchrotron"&$\sim 19\GeV$&$\pi^-$-Nucleus \\
\verb"Tevatron"&$1.96\TeV$&$\bar p p$ \\
\verb"LHC"&$7 -14\TeV$& $pp$\\
\br
\end{tabular}
\end{table}
\end{center}

\begin{figure}[h]
\includegraphics[width=22pc]{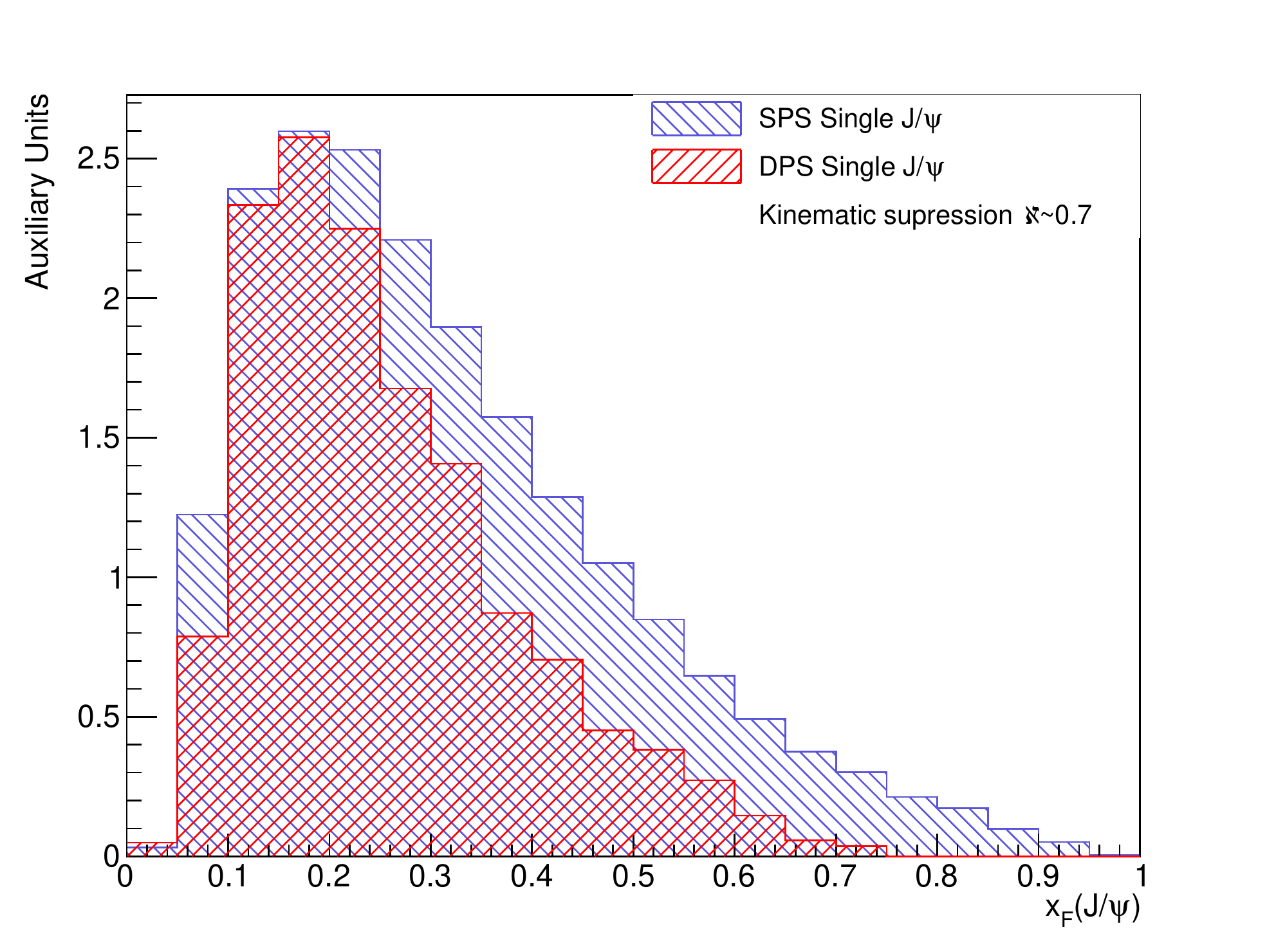}\hspace{2pc}%
\begin{minipage}[b]{14pc}\caption{\label{fig:aleph_DPS} 
Feynman-$x$ distribution of the production production cross section 
of the single $\Jp$ as independent process {\it i.e.} SPS and the respective 
distribution in case of DPS. Both distributions are obtained 
by using Pythia 8~\cite{Sjostrand:2007gs}. 
}
\end{minipage}
\end{figure}

In order to estimate the kinematic suppression at the COMPASS energy we
investigate the difference in the production of the single $\Jp$ in SPS and
DPS (cf.\ Fig.~\ref{fig:aleph_DPS}).
 As we can see, the $\Jp$'s from DPS are suppressed relatively to the $\Jp$'s 
from SPS. We can estimate the kinematic suppression factor as 
$\aleph \sim 0.7$. Accordingly, Eq.~(\ref{eq:DPS_general}) can be cast into
the form
\begin{equation}\label{eq:DPS_aleph}
\sigma_{\rm DPS}^{\pi^- A}(\Jp\Jp)= \frac{\aleph^2}{2}
\frac{\sigma(\Jp)^{\pi^- N}\sigma(\Jp)^{\pi^- N}}{\sigma_{\rm eff}^{\pi^- A}} \, .
\end{equation}

Utilizing the $\pi^-$ beam at $200\GeV/c$ with incident on hydrogen and
platinum targets, the NA3 experiment provided a single $\Jp$ cross section in
the $x_F^* > 0$ kinematic region,
$\sigma(\Jp) \times Br (\Jp \to \mu^+ \mu^-) = 6.3 \pm 0.9\nb/{\rm H_2}$ and
$\sigma(\Jp) \times Br (\Jp \to \mu^+ \mu^-) = 960 \pm 150\nb/{\rm Pt}$ per
nucleus~\cite{Badier:1983dg}. For a heavy nucleus like platinum or tungsten, 
$\sigma_{\rm eff}^{\pi^- A}$ is parametrized as
\begin{equation}
\frac{\sigma_{\rm eff}^{\pi^- N}} {\sigma_{\rm eff}^{\pi^- A}}
  \approx 3A \, (\sim 600).
\end{equation}
The value of $\sigma_{\rm eff}^{\pi^- N}$ is unknown. The value of 
$\sigma_{\rm eff}^{pp} \approx 5\mb$ is measured in double $J/\psi$ production 
(cf.\ Tab.~\ref{tab:sigma_eff}) and $\sigma_{\rm eff}^{\pi\pi}=41\mb$ is
calculated~\cite{Rinaldi:2018zng}. Comparing these values, we see that in
the pion--pion case the value of $\sigma_{\rm eff}$ is higher. Therefore, 
we can choose $\sigma_{\rm eff}^{pp} \approx 5\mb$ to obtain the upper limit
\begin{equation}
\sigma_{\rm DPS}^{\pi^- N}(\Jp\Jp) \sim 1\pb{\rm/nucleon}.
\end{equation}
\begin{center}
\begin{table}[h]
\caption{\label{tab:sigma_eff} $\sigma_{\rm eff}$ extracted from  
double $\Jp$ production data.}
\centering
\begin{tabular}{@{}*{7}{l}}
\br
Experiment & Energy & Colliding Mode & $\sigma_{\rm eff}$ (mb) \\
\mr
 D$\O$~\cite{Abazov:2014qba} & $1.96\TeV$ & $ p \bar p $ & $4.8\pm 2.5$  \\ 
 ATLAS~\cite{Aaboud:2016fzt}& $8\TeV$ & $pp$ & $6.3 \pm 1.9$ \\ 
 LHCb~\cite{Aaij:2016bqq} &  $13\TeV$ & $pp$ & $8.8-12.5$ \\ 
\br
\end{tabular}
\end{table}
\end{center}
\subsection{Single Parton Scattering}

\begin{figure}[h]
\begin{minipage}{14pc}
\includegraphics[width=17pc]{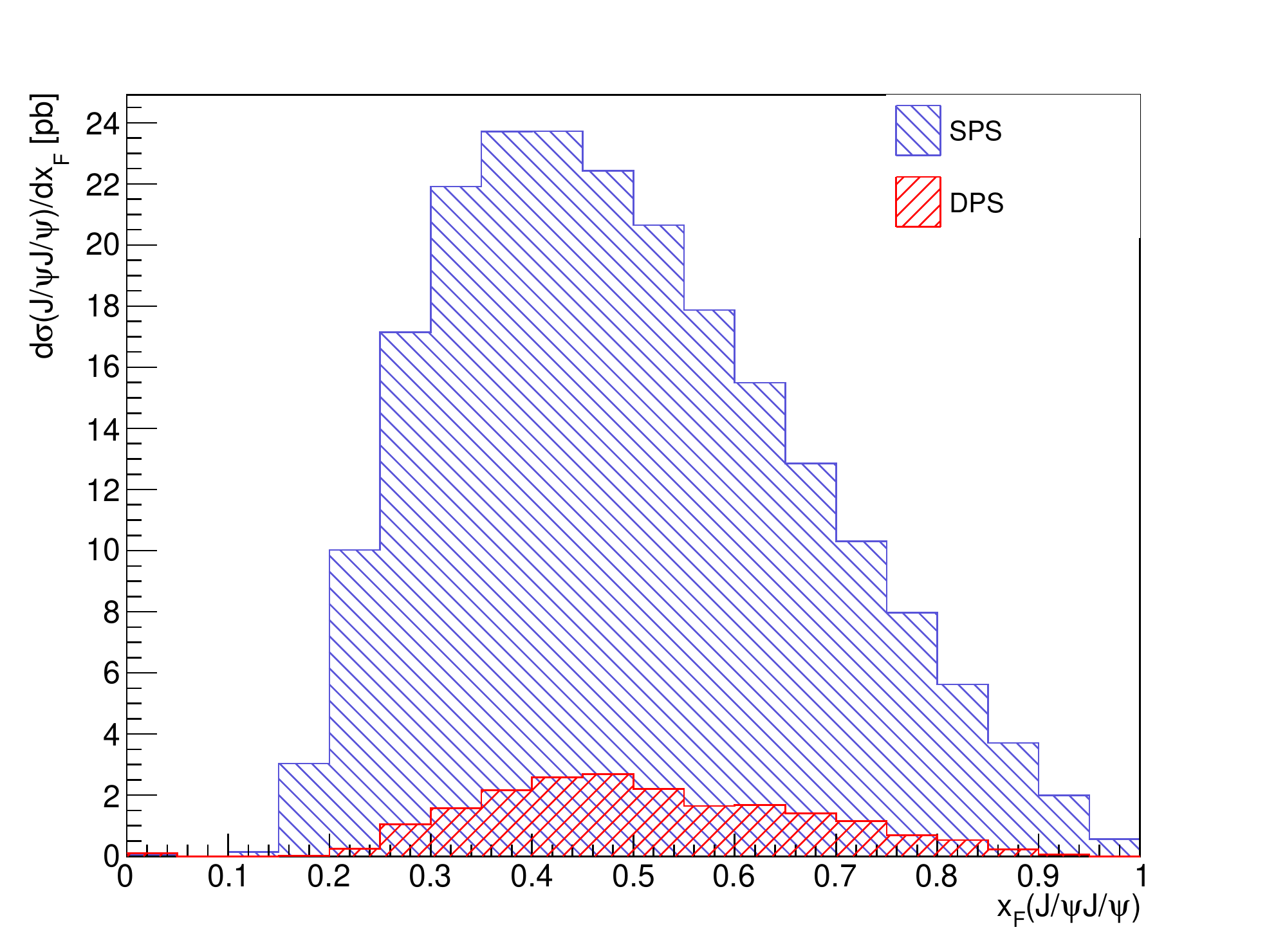}
\caption{\label{fig:Total_12} SPS and DPS 
double $\Jp$ production cross sections as Feynman-$x$ function for 
the total $\sigma_{\Jp\Jp} = 12\pb{\rm/nucleon}$. Both distributions are
obtained by using Pythia 8~\cite{Sjostrand:2007gs}.}
\end{minipage}\hspace{5pc}%
\begin{minipage}{14pc}
\includegraphics[width=17pc]{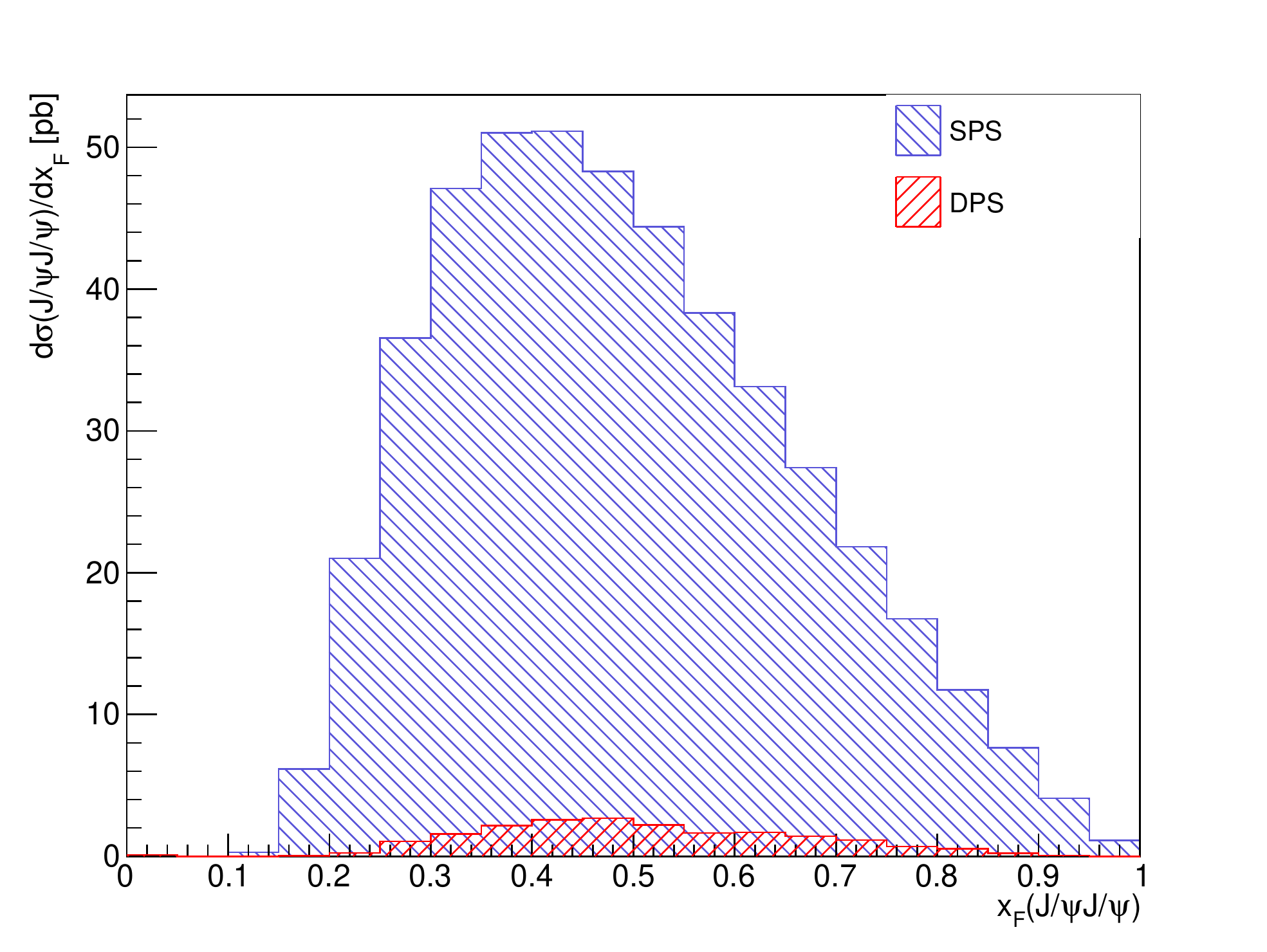}
\caption{\label{fig:Total_29} SPS and DPS 
double $\Jp$ production cross sections as Feynman-$x$ function for 
the total $\sigma_{\Jp\Jp} = 29\pb{\rm/nucleon}$. Both distributions are
obtained by using Pythia 8~\cite{Sjostrand:2007gs}.}
\end{minipage} 
\end{figure}

It is definitely interesting to investigated the role of DPS in the production 
of double $\Jp$ (cf.\ Figs.~\ref{fig:Total_12} and~\ref{fig:Total_29}).
Following calculations of the double $\Jp$ production cross section in SPS
by Ref.~\cite{Humpert:1983qt} we can find a ratio between the double $\Jp$ 
production cross section using a $\pi^-$ beam at NA3 and COMPASS energies:
\begin{equation}
 \sigma_{\Jp\Jp} (150\GeV/c) : \sigma_{\Jp\Jp} (190\GeV/c): \sigma_{\Jp\Jp} (280\GeV/c) 
 \approx 1 : 2.06 : 3.34.
\end{equation}
Using the mean values for the double $\Jp$ production cross section measured
by NA3 of $18 \pm 8 \pb$ and $30 \pm 10 \pb$ per nucleon at 150 and
$280\GeV/c$ with incident on a platinum target~\cite{Badier:1982ae} as
reference points, at $190\GeV/c$ we find
\begin{equation}
\sigma_{\Jp\Jp} \approx 12 - 29 \pb{\rm/nucleon}.
\end{equation}

For the Tevatron and LHC accelerators the leading production mechanism for
both DPS and SPS is gluon--gluon fusion. In contrast to that the leading
production mechanism of the double $\Jp$ in SPS at COMPASS energy with
$\pi^-$ beam is quark--antiquark annihilation~\cite{Humpert:1983qt}:
\begin{equation}
\sigma(q \bar q \to \Jp \Jp) / \sigma(gg \to \Jp \Jp) \approx 2.6.
\end{equation}
As the higher $x_F$ region is vanished out for the DPS (cf.\
Fig~\ref{fig:aleph_DPS}), the leading production mechanism for DPS is
gluon--gluon fusion.

\subsection{Feed-down effect in $\Jp$ production}

\begin{figure}[h]
\begin{minipage}{14pc}
\includegraphics[width=17pc]{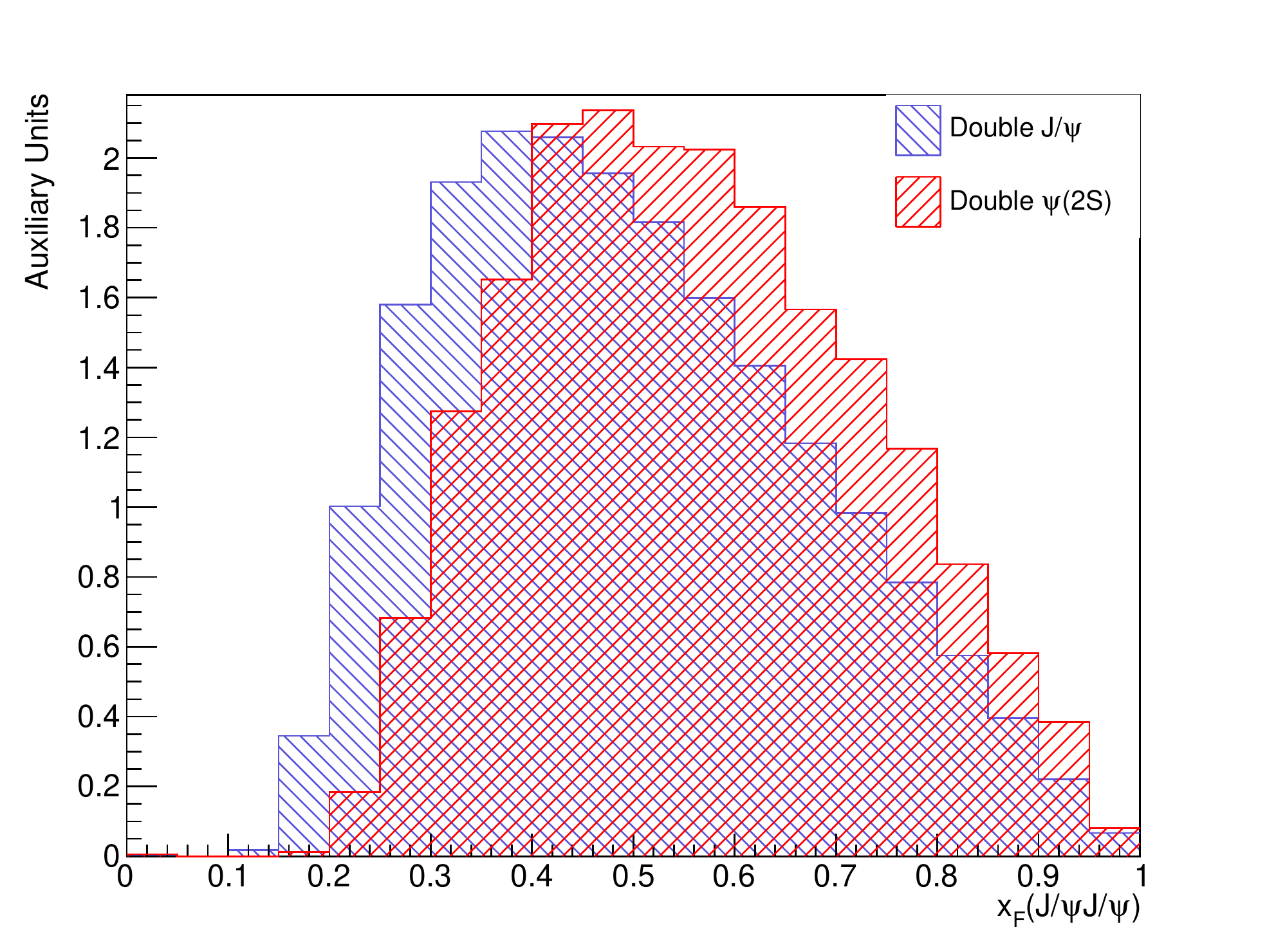}
\caption{\label{fig:feed_down_SPS} SPS  
double $\Jp$ production cross sections as Feynman-$x$ function for 
prompt-$\Jp$'s and from decays of $\psi(2S)$ in non-recoil regime. 
Both distributions are normalized to unity. The distributions are obtained 
by using Pythia 8~\cite{Sjostrand:2007gs}.}
\end{minipage}\hspace{5pc}%
\begin{minipage}{14pc}
\includegraphics[width=17pc]{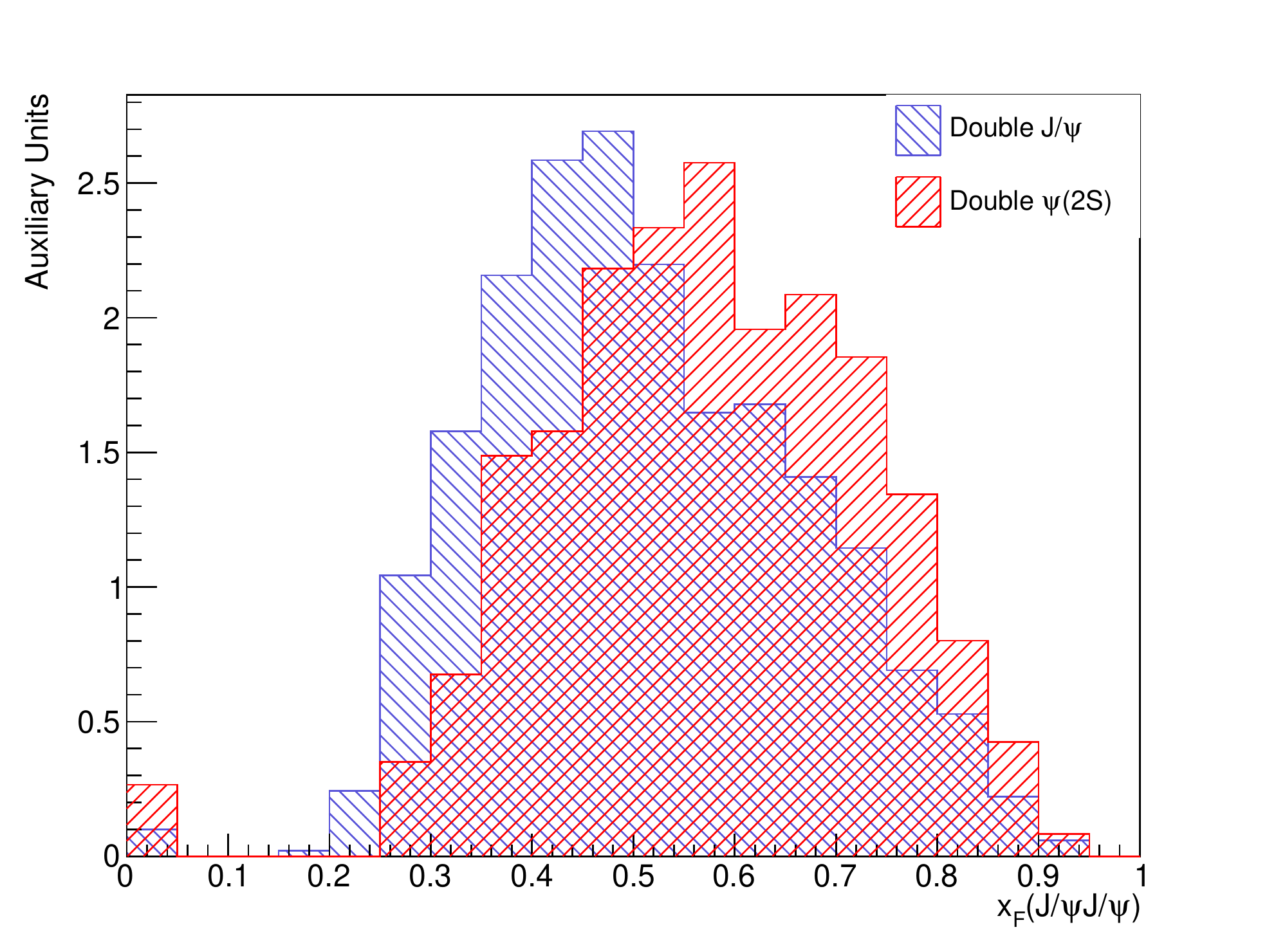}
\caption{\label{fig:feed_down_DPS} DPS  
double $\Jp$ production cross sections as Feynman-$x$ function for 
prompt-$\Jp$'s and from decays of $\psi(2S)$ in non-recoil regime. 
Both distributions are normalized to unity. The distributions are obtained 
by using Pythia 8~\cite{Sjostrand:2007gs}.}
\end{minipage} 
\end{figure}

For estimating of double $\Jp$ production cross sections for both SPS and DPS
we widely used measurements by the NA3 experiment. However, NA3 provided no
prompt-$\Jp$ data. From Tab.~\ref{tab:feed-down} we can learn the importance
of the overall $\Jp$ production cross section. Therefore, it is interesting to
investigate the possible impact of the feed-down effect on the Feynman-$x$
distributions (cf.\ Figs.~\ref{fig:feed_down_SPS} and~\ref{fig:feed_down_DPS}).
To make the difference in kinematics more visible, we use normalized
distributions instead of those re-weighted with respective feed-down fractions.
\begin{center}
\begin{table}[h]
\caption{\label{tab:feed-down} Cross sections for direct charmonium 
production in $\pi^-N$ collisions, normalized to the overall $\Jp$ production
cross section; branching fractions and feed-down fractions~\cite{Digal:2001ue}.}
\centering
\begin{tabular}{@{}*{7}{l}}
\br
State & Mass (GeV) & Decay mode ($Br$) & $f_{X} (\pi^-N)$ \\
\mr
 $J/\psi$ & 3.10 & -- & $0.57 \pm 0.03$ \\ 
 $\psi(2S)$ & 3.69 & $J/\psi + X$ (61\%) & $0.08 \pm 0.02$ \\ 
$\chi_{1c}(1P)$ & 3.51 & $J/\psi + \gamma$ (34\%) & $0.20 \pm 0.05$ \\ 
$\chi_{2c}(1P)$ & 3.56 & $J/\psi + \gamma$ (19\%) & $0.15 \pm 0.04$ \\
\br
\end{tabular}
\end{table}
\end{center}

\section{Conclusion}

Using both theoretical and experimental knowledge, we estimated the upper
limit for the Double Parton Scattering (DPS) cross section in double $\Jp$
production at the COMPASS detector using the CERN $\pi^-$ beam at $190\GeV/c$
to be $\sigma_{\rm DPS}^{\pi^- N}(\Jp\Jp) \sim 1\pb$ per nucleon. This value is
$\sim 3-8\%$ of the total double $\Jp$ production cross section. We found that
at the COMPASS energy the general formula for the predicted DPS cross section
should be modified by including an additional kinematic suppression factor
$\aleph^2 \sim 1/2$. The impact of the feed-down effect on the Feynman-$x$
distributions is also discussed.

\subsection*{Acknowledgments}

The author would like to thank Stefan Groote and Georgy Golovanov for
stimulating discussions, Andrei Gridin and Alexey Guskov for making this talk
possible, and Maxim Vtyurin for help with running the Monte-Carlo simulation.
At last but not at least the author would like to thanks Stefan Groote for
help and patience during the preparation of the manuscript. S.K. is supported
by the Estonian Research Council under Grant No.~PRG356.

\end{document}